\documentclass[prl,twocolumn,showpacs,floatfix,superscriptaddress]{revtex4-1}
\usepackage{graphicx}
\usepackage{hyperref}
\usepackage{bm}
\usepackage{dsfont}
\usepackage{rotating}
\usepackage{times}
\usepackage[utf8]{inputenc}
\usepackage{color}

\newcommand{\picscaling}{0.238}
\newcommand{\picbroad}{0.2618}
\newcommand{\picnarrow}{0.2142}

\begin{document}

\newcommand{\diff}{\mathrm{d}}

\newcommand{\pabl}[2]{\frac{\partial #1}{\partial #2}}
\newcommand{\abl}[2]{\frac{\diff #1}{\diff #2}}

\newcommand{\varphit}{\tilde{\varphi}}
\newcommand{\Mtilde}{\tilde{M}}
\newcommand{\Ktilde}{\tilde{K}}
\newcommand{\gammatilde}{\tilde{\gamma}}

\newcommand{\twovec}[2]{\left( \begin{array}{c} #1 \\ #2 \end{array}\right)}
\newcommand{\twotimestwo}[4]{\left( \begin{array}{cc} #1 & #2 \\ #3 & #4 \end{array}\right)}

\newcommand{\vekt}[1]{{\bm{#1}}}
\newcommand{\vect}[1]{\vekt{#1}}
\newcommand{\vektalpha}{\vekt{\alpha}}
\newcommand{\vekta}{\vekt{a}}
\newcommand{\vektr}{\vekt{r}}
\newcommand{\vektd}{\vekt{d}}
\newcommand{\vektx}{\vekt{x}}
\newcommand{\vektp}{\vekt{p}}
\newcommand{\NRNOmax}{N_{\mathrm{RNO}}}
\newcommand{\Ncyc}{N_{\mathrm{cyc}}}
\newcommand{\Vn}{V_{\mathrm{n}}}
\newcommand{\Vi}{V_{\mathrm{i}}}
\newcommand{\Hi}{H_{\mathrm{i}}}
\newcommand{\ee}{{\mathrm{ee}}}
\newcommand{\fee}{f_{\mathrm{ee}}}
\newcommand{\vektpc}{\vekt{p}_{\mathrm{c}}}
\newcommand{\vektE}{\vekt{E}}
\newcommand{\vekte}{\vekt{e}}
\newcommand{\vektA}{\vekt{A}}
\newcommand{\vektEhat}{\hat{\vekt{E}}}
\newcommand{\vektB}{\vekt{B}}
\newcommand{\vektv}{\vekt{v}}
\newcommand{\vektk}{\vekt{k}}
\newcommand{\vektkhat}{\hat{\vekt{k}}}
\newcommand{\reff}[1]{(\ref{#1})}

\newcommand{\calH}{{\cal H}}
\newcommand{\calP}{{\cal P}}
\newcommand{\calD}{{\cal D}}
\newcommand{\calh}{{\cal h}}
\newcommand{\calM}{{\cal M}}
\newcommand{\calK}{{\cal K}}

\newcommand{\Tr}{{\mathrm{Tr}}}
\newcommand{\Up}{U_{\mathrm{p}}}
\newcommand{\Ip}{I_{\mathrm{p}}}
\newcommand{\Tp}{T_{\mathrm{p}}}
\newcommand{\Ss}{S_{\vektp\mathrm{s}}}
\newcommand{\SIs}{S_{\vektp\Ip\mathrm{s}}}
\newcommand{\Cps}{C_{\vektp\mathrm{s}}}
\newcommand{\Cpnulls}{C_{\vektp_0\mathrm{s}}}
\newcommand{\SI}{S_{\vektp\Ip}}
\newcommand{\SIsnull}{S_{\vektp_0\Ip\mathrm{s}}}
\newcommand{\zt}{z_{\mathrm{t}}}
\newcommand{\ts}{t_{\vektp\mathrm{s}}}
\newcommand{\tsnull}{t_{\vektp_0\mathrm{s}}}
\newcommand{\tnulltilde}{\tilde{t}_0}
\newcommand{\omegap}{\omega_p}
\newcommand{\omegaMie}{\omega_M}
\newcommand{\imagi}{\mathrm{i}}
\newcommand{\eulere}{\mathrm{e}}

\newcommand{\halb}{\frac{1}{2}}

\newcommand{\phitilde}{\tilde{\phi}}
\newcommand{\alphatilde}{\tilde{\alpha}}
\newcommand{\phitildeperp}{\tilde{\phi}^{\perp}}

 \newcommand{\psihat}{{\hat{\psi}}}
 \newcommand{\psihatdag}{{\hat{\psi}^\dagger}}
\newcommand{\beq}{\begin{equation}}
\newcommand{\eeq}{\end{equation}}

\newcommand{\energy}{{\cal E}}

\newcommand{\Ehat}{\hat{E}}
\newcommand{\Ahat}{\hat{A}}
\newcommand{\ahat}{\hat{a}}
\newcommand{\ahatdag}{\hat{a}^\dagger}

\newcommand{\ket}[1]{\vert #1\rangle}
\newcommand{\bra}[1]{\langle#1\vert}
\newcommand{\braket}[2]{\langle #1 \vert #2 \rangle}

\newcommand{\makered}[1]{{\color{red} #1}}
\newcommand{\makegreen}[1]{{\color{green} #1}}
\newcommand{\makeblue}[1]{{\color{blue} #1}}

\renewcommand{\Re}{\,\mathrm{Re}\,}
\renewcommand{\Im}{\,\mathrm{Im}\,}

\newcommand{\varphic}{\varphi_{\mathrm{c}}}

\newcommand{\tini}{t_\mathrm{i}}
\newcommand{\tfinal}{t_\mathrm{f}}
\newcommand{\vxc}{v_\mathrm{xc}}
\newcommand{\vextop}{\hat{v}_\mathrm{ext}}
\newcommand{\VC}{V_\mathrm{c}}
\newcommand{\VHX}{V_\mathrm{Hx}}
\newcommand{\VHXC}{V_\mathrm{Hxc}}
\newcommand{\wop}{\hat{w}}
\newcommand{\Gammaevenodd}{\Gamma^\mathrm{even,odd}}
\newcommand{\Gammaeven}{\Gamma^\mathrm{even}}
\newcommand{\Gammaodd}{\Gamma^\mathrm{odd}}

\newcommand{\Hop}{\hat{H}}
\newcommand{\hop}{\hat{h}}
\newcommand{\Cop}{\hat{C}}
\newcommand{\HopKS}{\hat{H}_\mathrm{KS}}
\newcommand{\HKS}{H_\mathrm{KS}}
\newcommand{\Top}{\hat{T}}
\newcommand{\TopKS}{\hat{T}_\mathrm{KS}}
\newcommand{\VopKS}{\hat{V}_\mathrm{KS}}
\newcommand{\VKS}{{V}_\mathrm{KS}}
\newcommand{\vKS}{{v}_\mathrm{KS}}
\newcommand{\Ttildeop}{\hat{\tilde{T}}}
\newcommand{\Ttilde}{{\tilde{T}}}
\newcommand{\Vextop}{\hat{V}_{\mathrm{ext}}}
\newcommand{\Vext}{V_{\mathrm{ext}}}
\newcommand{\Vopee}{\hat{V}_{{ee}}}
\newcommand{\psiopdag}{\hat{\psi}^{\dagger}}
\newcommand{\psiop}{\hat{\psi}}
\newcommand{\vext}{v_{\mathrm{ext}}}
\newcommand{\Vee}{V_{ee}}
\newcommand{\nop}{\hat{n}}
\newcommand{\Uop}{\hat{U}}
\newcommand{\Wop}{\hat{W}}
\newcommand{\bop}{\hat{b}}
\newcommand{\bopdag}{\hat{b}^{\dagger}}
\newcommand{\qop}{\hat{q}}
\newcommand{\jop}{\hat{j\,}}
\newcommand{\vHxc}{v_{\mathrm{Hxc}}}
\newcommand{\vHx}{v_{\mathrm{Hx}}}
\newcommand{\vH}{v_{\mathrm{H}}}
\newcommand{\vc}{v_{\mathrm{c}}}
\newcommand{\xop}{\hat{x}}

\newcommand{\Wcmcm}{W/cm$^2$}

\newcommand{\varphiexact}{\varphi_{\mathrm{exact}}}

\newcommand{\fmathbox}[1]{\fbox{$\displaystyle #1$}}


\title{Laser-driven recollisions under the Coulomb barrier} 

\author{Th.~Keil}
\affiliation{Institut f\"ur Physik, Universit\"at Rostock, 18051 Rostock, Germany}

\author{S.V.~Popruzhenko}
\affiliation{National Research Nuclear University MEPhI, Kashirskoe shosse 31, 115409, Moscow, Russian Federation}

\author{D.~Bauer}
\affiliation{Institut f\"ur Physik, Universit\"at Rostock, 18051 Rostock, Germany}

\date{\today}

\begin{abstract} 
Photoelectron spectra obtained from the {\em ab initio} solution of the time-dependent Schrödinger equation can be in striking disagreement with predictions by the strong-field approximation (SFA) not only at low energy but also around twice the ponderomotive energy where the transition from the direct to the rescattered electrons is expected. In fact, the relative enhancement of the ionization probability compared to the SFA in this regime can be several orders of magnitude. We show  for which laser and target parameters such an enhancement occurs and for which the SFA prediction is {qualitatively} good. The enhancement is analyzed in terms of the Coulomb-corrected action along analytic quantum orbits in the complex-time plane, taking {soft recollisions under the Coulomb barrier into account. These  recollisions in complex time and space} {prevent a separation into sub-barrier motion up to the ``tunnel exit'' and subsequent classical dynamics. Instead, the entire quantum path up to the detector determines the ionization probability.}

\end{abstract}
\maketitle

Various surprises in strong-field photoelectron spectra (PES) have been found recently, especially at low energies and for long wavelengths \cite{Wolter2015}.  Here, by ``surprise'' we mean ``in disagreement with the strong-field approximation'' (SFA) \cite{Keldysh1964,Faisal1973,Reiss1980} or tunneling theories~\cite{Popov2004}.
The SFA is the theorists' work horse for intense-laser ionization problems and has been particularly insightful thanks to its formulation in terms of quantum orbits \cite{Kopold2000,Salieres2001,Milosevic2006,Popruzhenko2014a}. However, in plain SFA, for the so-called direct SFA matrix element, the Coulomb interaction of the outgoing  photoelectron with its parent ion is neglected.
Laser-driven recollisions can be taken into account in an extended SFA via a Born-like rescattering matrix element \cite{Reiss1980,Lohr97,Milosevic1998}, leading to good agreement with experimental results or {\em ab initio} solutions of the time-dependent Schrödinger equation (TDSE) for high photoelectron momenta or emission angles  where the direct SFA matrix element alone yields exponentially small  probabilities \cite{Moeller14}. 
The strongest discrepancies between SFA and TDSE or experiment were found in the low \cite{Blaga2009,Faisal2009}, very low \cite{Quan2009,Wu2012}, and zero  energy \cite{Dura2013,Wolter2015} regime where surprisingly high ionization probabilities for particular final electron momenta were observed.  
All these low-energy structures were found to originate from soft (multiple) laser-driven  recollisions~\cite{LiuHatsag10,Kaestner12}. Their locations in momentum space are even encoded in the rescattering SFA matrix element \cite{Moeller14,Milosevic2014} but they do not stick out probability-wise without taking interaction with the parent ion into account.  

In this Letter, we examine the energy regime around twice the ponderomotive energy. The  ponderomotive energy $\Up$ is the time-averaged kinetic energy of a free electron in a laser field. 
For a linearly polarized laser field  of the form $E(t)=\hat E \cos\omega t$ (in dipole approximation) the velocity $v$ of a classical electron emitted at time $t_0$ with $v(t_0)=0$ is, at $t\to\infty$, when the laser is off, $v=(\hat E/\omega)\sin\omega t_0$ (atomic units are used unless noted otherwise). The classically highest kinetic energy $\hat E^2/(2\omega^2)=2\Up$ thus results for ionization times $t_0$  where $|\sin\omega t_0|=1$. Since at such times $E(t)=0$, tunneling rate formulas \cite{Popov2004}  will give zero weight to such electrons. However, as shown in Fig.~\ref{fig1} below, $2\Up$-photoelectrons are clearly observed in spectra obtained solving numerically the TDSE. The SFA, as a quantum mechanical approach, has no abrupt cut-off. In fact, the SFA has not even a  cut-off-like feature around $2\Up$ such as a change in slope. Apart from the well understood intra and inter-cycle interferences \cite{Arbo10}, SFA spectra obtained using the direct matrix element alone just roll-off featureless. At some photoelectron energy---typically around (2--4)$\Up$---the rescattering SFA matrix element takes over, up to the rescattering cut-off around $10\Up$ from where on  the photoelectron yield ceases quickly with increasing energy. The SFA rescattering matrix element could be made very  strong between zero and $2\Up$ energy by reducing the screening in the model potential. In fact, for potentials with a Coulomb tail, the rescattering probability for energies between zero and $2\Up$ can be comparable to the probability for direct ionization. The SFA rescattering matrix element then generates an overpronounced, step-like plateau up to $2\Up$, which is in even worse agreement with the TDSE spectrum than the probability from the direct SFA matrix element alone. 

The main purpose of this Letter is {threefold}: (i) to predict for which laser and target parameters the direct SFA spectrum is in good or bad agreement with the TDSE result concerning the yield around $2\Up$, (ii)  {to highlight} the mechanism that leads to an enhancement in the yield employing Coulomb-corrected quantum orbits in complex space and time, {and (iii) to show that the notion of a ``tunnel exit'' after which the electron dynamics can be treated classically is erroneous in this case.    In fact, i}t turns out that none of the commonly applied Coulomb corrections of orbits starting from the tunnel exit in real space and time (incl.\ our own used in Ref.~\cite{Yan2010}) makes up for the enhanced yield around $2\Up$, simply because the ionization probability is determined once the electron arrived at the tunnel exit in such approaches.  Instead, the proper incorporation of soft recollisions via navigating the quantum orbits through Coulomb-generated branch cuts in the complex-time plane is required, as was pointed out in Refs.~\cite{Popruzhenko2014,Pisanty2016}.

A prominent example for a distinct $2\Up$-plateau in the photoelectron yield can be found in Fig.~3 of Ref.\ \cite{Huismans2011}. In this experiment, a short,
intense 7-micron laser pulse ionized Xe atoms prepared in the 6s excited, metastable state of ionization potential $\Ip=0.14$.  In Fig.\ \ref{fig1}(a)  a momentum-resolved PES in laser polarization direction from the numerical solution of the TDSE  is shown for the laser parameters of Ref. \cite{Huismans2011}. In order to reproduce $\Ip$ in single-active-electron approximation the 2s state in an effective potential $V_\mathrm{eff}(r)=-[Z + (Z_\mathrm{full}-Z)\eulere^{-r/r_\mathrm{s}})]/r$ was used as the initial state, where $Z=1$ and $Z_\mathrm{full}=54$  are the asymptotic ion charges as $r\to\infty$ and $r\to 0$, respectively, and a screening length $r_\mathrm{s} = 0.026$ was tuned to match $\Ip$.
As the SFA cannot be expected to give correct ionization probabilities in absolute numbers, we are free to shift the corresponding SFA spectrum vertically. We may adjust it to match the TDSE result in the rescattering-plateau region or at low energies. In either case we observe a discrepancy between TDSE and SFA of more than an order of magnitude for energies between $\Up$ and $4\Up$. The prediction of a ``simple man's model'' (SMM, {see, e.g., \cite{CorkumSMT}}) where each ionization time $t_0$ introduced above is weighted by a tunneling rate is included in Fig.~\ref{fig1}(a). The SMM spectrum drops $\sim\exp[-\frac{2}{3F} (1-q^2)^{-1/2}]$ where $F=\hat E/(2\Ip)^{3/2}$ is the reduced electric field and $q=p/\sqrt{4\Up}$, i.e., even faster with increasing momentum than the SFA, as it approaches zero at the cut-off momentum $p=\sqrt{4\Up}$.

\begin{figure}
  \centering
  \includegraphics[width=\picscaling\textwidth]{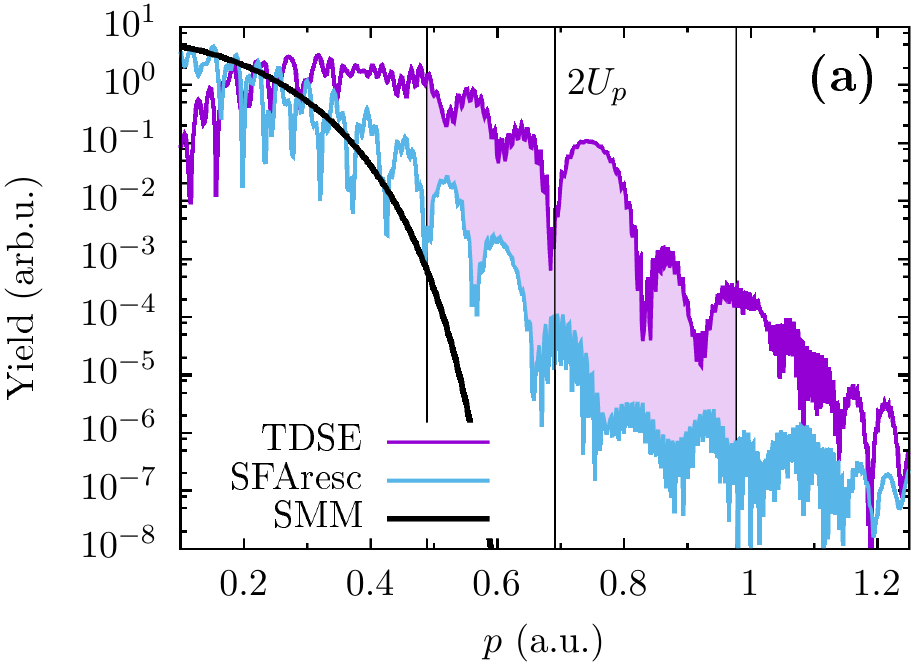}
  \includegraphics[width=\picscaling\textwidth]{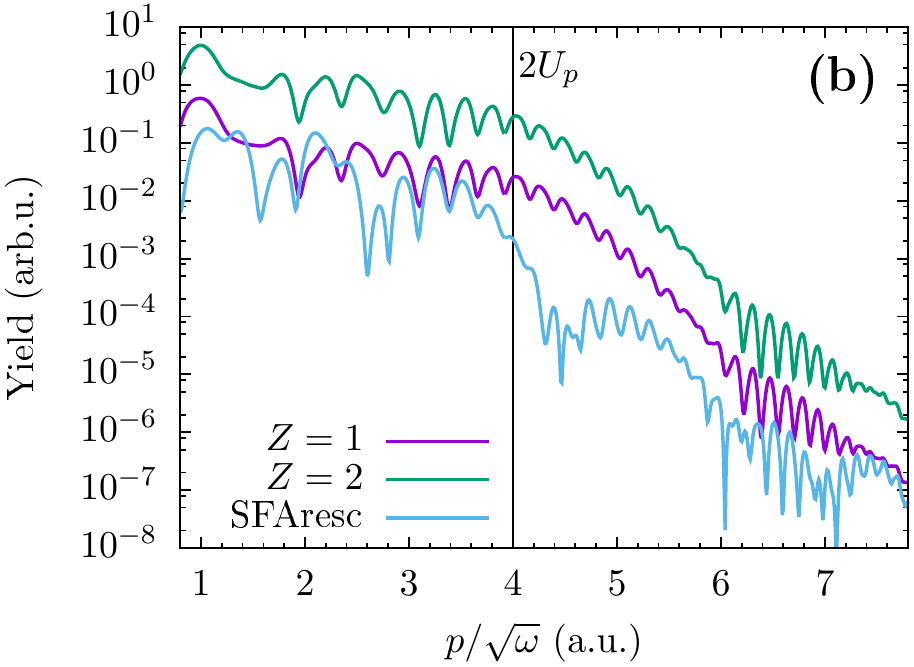}
  \includegraphics[width=\picscaling\textwidth]{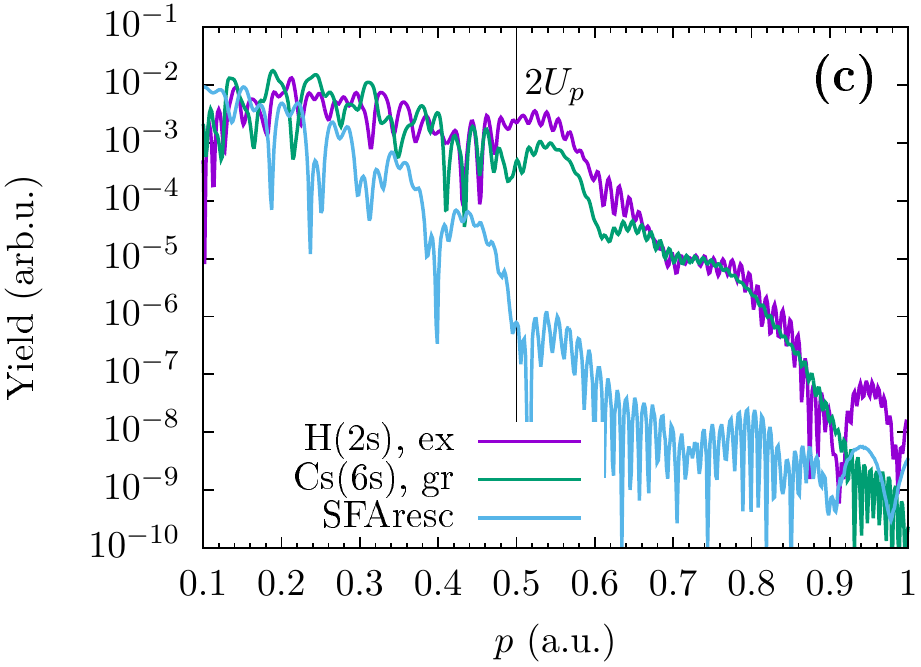}
  \includegraphics[width=\picscaling\textwidth]{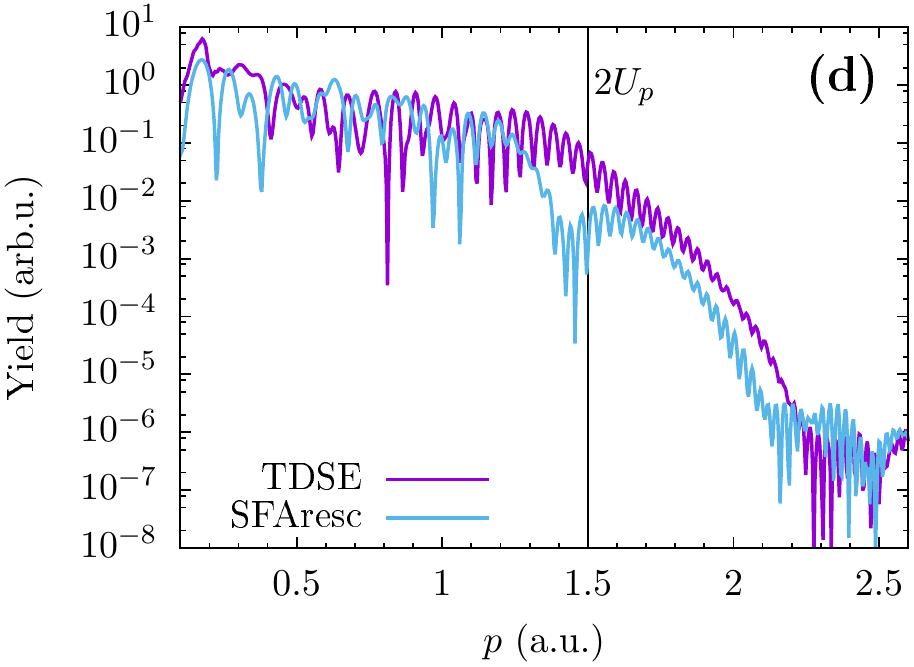}
\caption{\label{fig1}(Color online). TDSE spectra around the $2\Up$ cut-off compared to
plain-SFA spectra. (a) same $\Ip=0.14$ (3.8\,eV) and laser parameters as in Ref.\ \cite{Huismans2011}, i.e., $\omega = 0.0065$ (7 microns), $\hat E=0.0045$ ($I=7.1\times 10^{11}$\,\Wcmcm), $\alpha_\mathrm{C}= 12.4$ (asymptotic charge $Z=1$), $\alpha_\mathrm{L}= 8.6$, using a $6$-cycle $\sin^2$-shaped pulse envelope.  The shaded area between TDSE and SFA spectrum highlights the order-of-magnitude discrepancy. Vertical lines indicate 1, 2, and 4$\Up$. The SMM prediction is plotted bold black. Panel (b) shows two TDSE PES vs scaled momentum, both for the groundstate as initial state  and both with $\alpha_\mathrm{C}=\alpha_\mathrm{L}=4$ but one for $Z=1$, $\omega=0.0625$, $\hat E=0.0625$ (purple) and the other for $Z=2$, $\omega=0.25$, $\hat E=0.5$ (green, shifted for better comparison). The SFA result is included (cyan). Panel (c) shows two TDSE PES for similar $\Ip$ realized via an excited state (purple) and an effective potential (green), respectively (see text). In both cases  $\alpha_\mathrm{C}= 11.3$ and $\alpha_\mathrm{L}= 5.7$. The corresponding SFA spectrum is in striking disagreement. Panel (d) shows TDSE and SFA PES in good agreement for H(1s) as in (b) but for $\alpha_\mathrm{L}=6$. 
}
\end{figure}

In order to determine the relevant parameters that govern the (dis)agreement  between TDSE and SFA let us consider first the TDSE for the wavefunction $\Psi(\vektr,t)$ describing the electron in a hydrogen-like ion and a linearly polarized laser field. Expressed in dimensionless time $\tau=\omega t$ and position $\bar\vektr = \sqrt{\omega}\vektr$  (in SI units see \footnote{The scaling in SI units is $\tau=\omega t$
 and $\bm{\bar{r}} = \sqrt{m\omega/\hbar}~\vektr$.}) the TDSE reads
\beq \imagi\pabl{}{\tau}\bar\Psi = \left[-\halb \Delta_{\bar\vektr} -\frac{Z}{\sqrt{\omega}\bar r} +\frac{\hat E}{\omega^{3/2}} \bar z f(\tau)\cos\tau \right]\bar\Psi \eeq
with $\bar\Psi=\Psi[\vektr(\bar\vektr),t(\tau)]$ and $f(t)\leq 1$ a dimensionless envelope function. The scaling was chosen such that the left hand side (energy) and the kinetic energy (first term on the right hand side) keep their form, i.e., do not acquire an extra prefactor. Then the Coulomb potential scales with a factor $\alpha_\mathrm{C}=Z/\sqrt{\omega}$ and the laser-field with $\alpha_\mathrm{L}=\hat E/\omega^{3/2}=\sqrt{z_\mathrm{F}}$, where $z_\mathrm{F}$ is the strong-field parameter {\cite{Reiss1980}}. For hydrogen-like ions $\Ip=Z^2/(2n^2)$ where $n=Z/\sqrt{2\Ip}$ is the principal quantum number. For ground states  $n=1$ in hydrogen-like ions,  the ratio $\alpha_\mathrm{C}/\alpha_\mathrm{L}$ thus equals the Keldysh parameter $\gamma=\sqrt{\Ip/(2\Up)}$ \cite{Keldysh1964}.

For fixed $\alpha_\mathrm{C}$, $\alpha_\mathrm{L}$, and initial-state quantum numbers but otherwise different laser and target parameters the same PES are obtained when plotted vs the  dimensionless momentum $\vektp/\sqrt{\omega}$. This is demonstrated in Fig.~\ref{fig1}(b) for $\alpha_\mathrm{C}=\alpha_\mathrm{L}=4$ (the other parameters are given in the figure caption).

The plain SFA is known to depend on only two dimensionless parameters as well, e.g., any pair of the set $\gamma$, $z_\mathrm{F}$, the reduced electric field $F$, and the multiquantum parameter $K_0=\Ip/\omega$. As a consequence, the plain SFA predicts the same scaled spectrum too for the two cases in  Fig.~\ref{fig1}(b) (plotted only once). The  intra-cycle interference pattern is overpronounced in the SFA PES, which can lead to orders-of-magnitude discrepancies with the TDSE around certain photoelectron momenta (here $p/\sqrt{\omega} \simeq 4.4$). Apart from that  the overall slope of the SFA PES around  $2\Up$ is in good agreement with the TDSE.

Our simple scaling of the TDSE with only two dimensionless parameters  was only possible for pure, hydrogen-like Coulomb potentials. In practice one often wants to analyze experimental results for certain many-electron targets where $\Ip$ and the asymptotic charge $Z$ are prescribed. In single-active-electron TDSE simulations one takes these conditions into account by tuning the screening of an effective potential $V_\mathrm{eff}$ (e.g., as introduced above) to match $\Ip$. Writing $\alpha_\mathrm{C}$ in terms of the ionization potential for hydrogen-like ions $\Ip=Z^2/(2 n^2)$ we obtain $\alpha_\mathrm{C}=n\sqrt{2\Ip}/\sqrt{\omega}$, i.e., $\alpha_\mathrm{C}^2=2n^2K_0$.  Fixing $\Ip$ and $Z$ independently is like changing the principal quantum number $n$ of the initial state in a hydrogen-like ion accordingly. This variable effective principal quantum number is a third dimensionless parameter that comes into play when working with non-hydrogen-like effective potentials (or when starting from different initial states \footnote{In general, PES also depend on the other quantum numbers such as orbital angular momentum $l$ and magnetic quantum number $m$ but this dependence is not related to the yield enhancement discussed in this work.}).

 Figure \ref{fig1}(c) shows a TDSE PES for $\alpha_\mathrm{C}= 11.3$ and $\alpha_\mathrm{L}=5.7$ for $Z=1$, $n=2$ (2s-state, $\Ip=0.125$), $\omega=0.0078$, and $\hat E=0.0039$. The second TDSE spectrum in this panel was calculated for the groundstate of $\Ip=0.143$ (cesium) using the effective potential introduced above with $Z_\mathrm{full}=0.1$ and $r_\mathrm{s} = 3.13$ to match $\Ip$.
The TDSE spectra are very similar, showing that mainly $\Ip$ and the asymptotic $-Z/r$ matter for the overall shape of the PES. Note that the TDSE spectra are rather flat up to $2\Up$, in striking disagreement with the SFA \footnote{We found in Fig.~1 of Ref.~\cite{Milosevic1998} a flat PES plateau up to  $2\Up$, although without discussion. The PES there was calculated using a Coulomb-Volkov ansatz.} for an s-state with $\Ip=0.125$ despite the fact that the Keldysh parameter is still $\gamma= 1$, as in Fig.~\ref{fig1}(b) where the agreement with the SFA is good. This shows that the Keldysh parameter alone is not sufficient to characterize the importance of Coulomb corrections. 

Figure \ref{fig1}(d) shows an example where both TDSE and SFA PES display a rather flat $2\Up$ plateau and a slow roll-off down to the rescattering plateau.  The same parameters as for the H(1s) case in Fig.~\ref{fig1}(b) were used, apart from the increased field amplitude $\hat E=0.09375$, i.e., $\alpha_\mathrm{L}=6$. 

The TDSE solutions yield little insight as to how the binding potential modifies PES. The SFA, on the other hand, can be interpreted (and evaluated) using quantum orbits in complex spacetime. Starting point is the direct SFA matrix element in saddle-point approximation \cite{Milosevic2006,Popruzhenko2014a}
$ M_\vektp^\mathrm{(SFA)} = \sum_\alpha f_{\Psi_0}(\vektp,t_{0\alpha})\eulere^{\imagi S_{\vektp,\Ip}(t_{0\alpha})}$
where the sum runs over all complex solutions $t_{0\alpha}$ of the saddle-point equation $[\vektp+\vektA(t_{0\alpha})]^2/2=-\Ip$
for a given final photoelectron momentum $\vektp$. The vector potential $\vektA(t)$ is related to the electric field according $\vektE(t)=-\partial_t\vektA(t)$. 
The phase $S_{\vektp,\Ip}$ is the classical action along an electron orbit in the laser field $\vektA(t)$, 
\beq S_{\vektp,\Ip}(t_{0\alpha}) = -\int_{t_{0\alpha}}^{\Tp} \left\{ \halb [\vektp+\vektA(t)]^2 + \Ip \right\}\diff t, \label{eq:actionI} \eeq
 but evaluated for complex times, making the orbit ``quantum.'' At the upper integration limit  $\Tp$ the laser pulse is off again. For $t>\Tp$ all interfering trajectories with momentum $\vektp$ evolve equally and the propagation to the detector at $t\rightarrow\infty$ yields a constant phase factor only. The pre-exponential factor $f_{\Psi_0}(\vektp,t_{0\alpha})$ is proportional to the matrix element $\bra{\vektp+\vektA(t_{0\alpha})}E(t_{0\alpha})z\ket{\Psi_0}$ and thus acts like a  form factor that depends on the initial state $\Psi_0$.

We Coulomb-correct the plain SFA along the lines of Refs.~\cite{Popruzhenko2008,Yan2010,Popruzhenko2014a} (Coulomb-corrected SFA) or \cite{Torlina2012,Kaushal2013} (analytical $R$-matrix method). When applied in the perturbative regime both methods consider the change in the action due to the binding potential by integrating $V[\vektr(t)]$ along the unperturbed, complex plain-SFA quantum orbit 
\beq \vektr(t)=\vektp(t-t_{0\alpha})+\int_{t_{0\alpha}}^{t} \vektA(t')\,\diff t',\label{eq:orbit}\eeq
leading to
\beq \Delta S_\mathrm{CC}(t_{0\alpha}) = - \int_{t_{0\alpha}}^{\Tp} V[\vektr(t)]\,\diff t. \label{eq:DeltaScc}\eeq
The singularity of $V(\vektr)=-Z/r$ at the initial position $\vektr(t_{0\alpha})=\vekt{0}$ is regularized through matching to the field-free initial-state wavefunction \cite{Popov2004,Popruzhenko2008,Popruzhenko2014a}. 

\begin{figure}
  \includegraphics[width=\picbroad\textwidth]{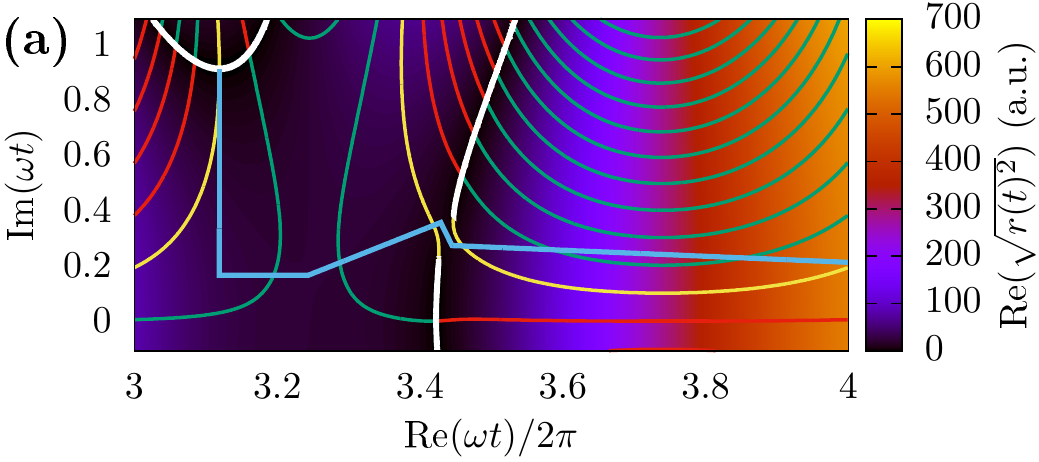}
  \includegraphics[width=\picnarrow\textwidth]{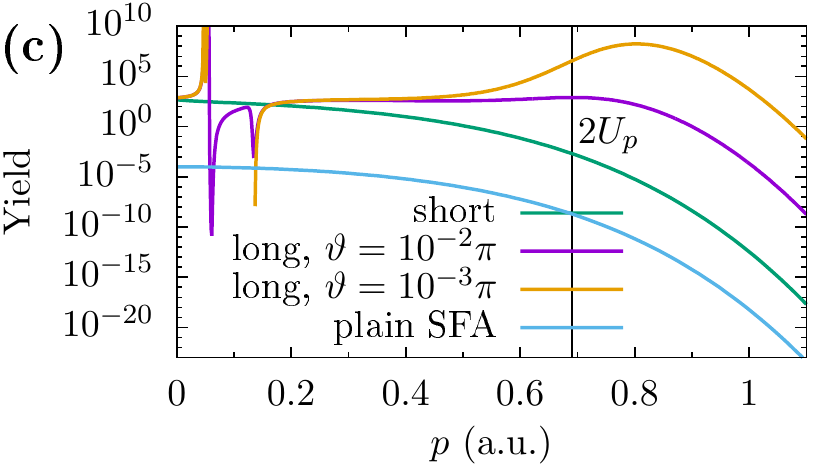}
  \includegraphics[width=\picbroad\textwidth]{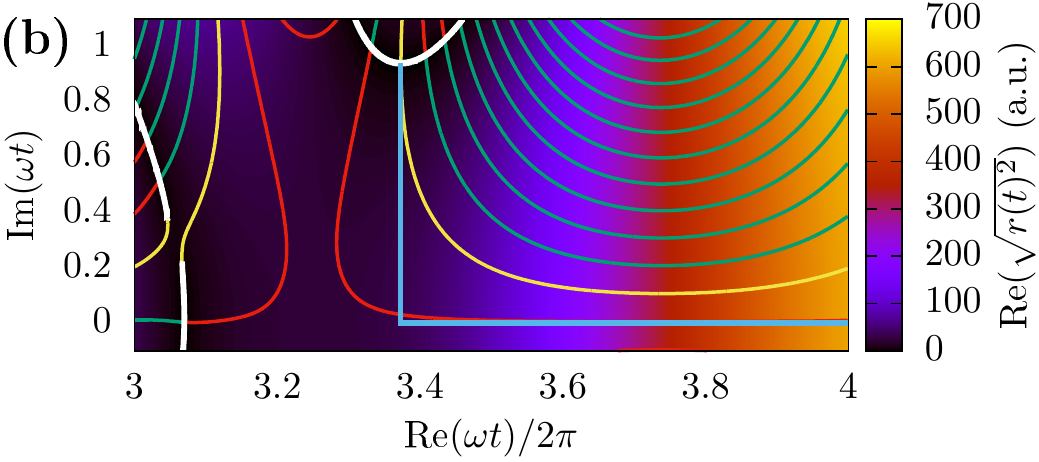}
  \includegraphics[width=\picnarrow\textwidth]{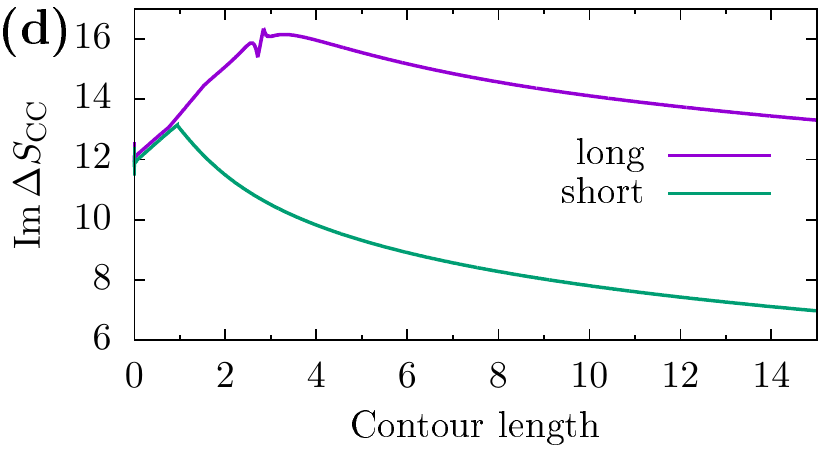}
\caption{\label{fig2}(Color online). How long and short orbits are affected by the Coulomb correction [same parameters as in Fig.~\ref{fig1}(a)]. Panels (a) and (b) show integration contours (cyan) in the complex-time plane for a long and a short orbit at the cut-off momentum, respectively. The color scale indicates $\Re \sqrt{\vektr^2(t)}$, the colored lines constant  $\Im \sqrt{\vektr^2(t)}$ (red $<0$, yellow $=0$, green $>0$). White lines indicate branch cuts starting at branch points. For the  short trajectory in (b) the integration contour can be chosen in the standard way {because the recollision branch points and cuts are far away from the real axis, i.e., outside the complex-time domain shown in (b)}.  For the  long trajectory in (a) the integration path needs to be deformed in order to navigate through the branch cut gate.  Panel (c) shows partial spectra from long and short trajectories for different angles $\vartheta$ and a plain SFA spectrum (calculated with only one trajectory per momentum to inhibit interferences). The angular dependence of the short-trajectory spectrum is negligible. Panel (d) displays the trajectory weights $\Im\Delta S_{\mathrm{CC}}$, measured along the integration contours in (a) and (b). Note that every kink in (d) is related to a change in the integration direction in (a), (b).}
\end{figure}

Without Coulomb correction the action \reff{eq:actionI} can be evaluated along any integration path in the complex-$t$ plane from complex $t_{0\alpha}$ to real $\Tp$, thanks to the analyticity of the integrand. When including the Coulomb-correction in \reff{eq:DeltaScc} for $V(\vektr)=-Z/\sqrt{\vektr^2}$ one has to take into account the branch points $\vektr^2=0$ and branch cuts originating from them along, e.g., $\Im \vektr^2=0$, $\Re \vektr^2 < 0$, and make sure that the integration path remains on a fixed Riemann sheet. The topology of branch cuts has been analyzed in Refs.~\cite{Popruzhenko2014,Pisanty2016}. Proper integration pathways and their effect on PES have been investigated  in \cite{Pisanty2016} to reveal the origin of low-energy structures. To identify the mechanism behind the enhanced ionization yield around $2\Up$ we performed a similar analysis. First, we calculated (for each final momentum $\vektp$ of interest) the positions of the branch points determined by the plain-SFA quantum orbit \reff{eq:orbit} for $\vektr^2(t)=0$. The roots of this equation come in pairs, see Fig.~\ref{fig2}(a). For so-called short orbits that do not return closely to the core, no obstacles in the form of branch cuts are put in the way of the ``standard'' integration path, which is down to the real-time axis at $t=\Re t_{0\alpha}$ (interpreted as the ``tunnel exit'' $\Re \vektr(\Re t_{0\alpha})$ in position space) and then along the real axis to $t=\Tp$, as in Fig.~\ref{fig2}(b). For so-called long orbits with close approaches to the core the pair of branch points are closer together, leaving a narrow gap only at finite imaginary-time values to navigate the integration path through, see Fig.~\ref{fig2}(a). 
As a consequence, the standard integration path is blocked by a branch cut. The algorithm used to find an analytic integration path first determines all returns on the real-time axis, then computes the corresponding branch point pair or ``gate'' as solutions of $\vektr^2(t)=0$. These are used to divide solutions of $\vektr(t) \cdot \vektv(t) = 0$---the principal waypoints---into times of closest approach (which represent the center of a gate) and intermediate turning points (corresponding to classical turning points) {\cite{Pisanty2016}}. These waypoints are then used to construct a valid, analytic  integration path passing perpendicularly through the gates.

The effect of the Coulomb-correction to the action \reff{eq:DeltaScc}  for the dominating orbits is shown in Fig.~\ref{fig2}(c). 
For the plain SFA the contributions
of short and long trajectories are equal. The
Coulomb correction for short trajectories increases the ionization probability uniformly. This
is consistent with previous Coulomb-corrected total ionization
rates \cite{Popov2004} which are found by calculating   the Coulomb correction along the unperturbed, dominant $\vektp =\vekt{0}$-trajectory analytically.
For long trajectories the spectrum is qualitatively
modified. For small momenta low-energy structures
are recovered [purple curve in Fig.~\ref{fig2}(c)], as in Ref.~\cite{Pisanty2016}. Around $2\Up$ the yield is significantly increased compared to 
the short-orbit contribution, generating a plateau in the spectrum. The magnitude of this increase depends
on the angle $\vartheta=\arccos(p_\parallel/p)$. For $\vartheta=0$, i.e., in polarization direction, this correction diverges logarithmically, and for very small $\vartheta$ it becomes large [see $\vartheta=10^{-3}\pi$ in Fig.~\ref{fig2}(c), yellow]. This is not surprising, as our perturbative Coulomb correction in its current form is not applicable for too close branch points \cite{Popruzhenko2014a} where a correct calculation requires a matching of the integral  \reff{eq:DeltaScc} with the phase of the atomic wave function. This will be the subject of future work.  For  $\vartheta=10^{-2}\pi$ in Fig.~\ref{fig2}(c) (purple)  our method recovers the plateau shape found in the TDSE spectra.  {We checked that the yield around $2\Up$ quickly drops with further increasing angle $\vartheta$, in agreement with Fig.~3A and 3B in \cite{Huismans2011}.} The imaginary part of the Coulomb action \reff{eq:DeltaScc} governs the weight of a quantum orbit's contribution to the photoelectron yield for a given  final momentum $\vektp$.  This quantity is shown in Fig.~\ref{fig2}(d) 
 as it is accumulated along the pathways shown in Fig.~\ref{fig2}(a) and (b). For both long and short orbit $\Im\Delta S_\mathrm{CC}$ first increases (corresponding to the integration path vertically downwards). Then $\Im\Delta S_\mathrm{CC}$ for the short trajectory decreases during the path along the real-time axis (note that it changes at all because $\sqrt{r^2}$ still has an imaginary part there), reaching asymptotically a constant value, determining the orbit's contribution to the yield.  For the long orbit $\Im\Delta S_\mathrm{CC}$ continues to increase until the branch-point gate is passed, and the path can be directed towards the real-time axis. The net enhancement in the imaginary part of the action increases the weight of this long trajectory. The fact that $t$ necessarily remains complex until it has passed the final recollision gate implies that the ultimate weight of a particular quantum orbit is not yet determined at $\Re t_{0\alpha}$. Of course, all these arguments belong to the realm of physical interpretation of mathematics  rather than of quantum mechanical observables. However, they are useful to highlight the insignificance of the tunneling exit.
Depending on the integration path the tunnel exit could be as well at any other point in space. 

In summary, orders-of-magnitude disagreements between SFA and TDSE PES can occur at intermediate electron energies around $2\Up$. 
The greater the ratio $\alpha_\mathrm{C}/\alpha_\mathrm{L}=n\gamma$   the bigger the expected disagreement  between SFA and TDSE PES. Indeed, for the four examples of Fig.~\ref{fig1} these ratios are (a) $1.44$ (disagreement), (b) $1.0$ (agreement), (c) $2.0$ (disagreement), and (d) $0.66$ (agreement). {The ratio is $<1$ for noble gas atoms in the ground state and typical near-infrared laser wavelengths and intensities. We therefore propose to check the $2\Up$ enhancement experimentally by using alkali atoms as targets  (see the cesium example in Fig.~\ref{fig1}c).}  Based on complex, plain-SFA quantum orbits we implemented a perturbative Coulomb correction to the action, taking branch points and cuts in the complex-time plane into account. We showed that the enhanced photoelectron  yield around $2\Up$ originates from those parts of the integration path that pass through gates made up by pairs of branch points. Physically, these parts of the integration paths correspond to soft recollisions. {It is crucial that these recollisions take place in complex time and therefore affect the ionization probability.}
 Coulomb corrections relying on adiabatic tunneling probabilities determined already at the ``tunnel exit'' could modify the yield only by focusing classical trajectories towards the corresponding momentum range, which can hardly explain an orders-of-magnitude increase over an extended energy interval around $2\Up$. In contrast, the plain SFA is nonadiabatic in that sense, and its analytic quantum orbits allow for arbitrarily early or late arrivals at the real-time axis. The analytic Coulomb correction used in this Letter allows the ultimate approach of the real-time axis only after the last recollision. The laser-induced recollisions are thus an integral part of the emission process because  the entire quantum orbit determines the yield {nonlocally}. 

\medskip

TK and DB acknowledge support through the SFB 652 of the German Science Foundation (DFG). SP acknowledges support from  the MEPhI Academic Excellence Project (contract No.\ 02.a03.21.0005, 27.08.2013) and the Russian Foundation for Basic Research (project No.\ 16-02-00936).

\bibliography{bibliography}

\begin{thebibliography}{32}%
\makeatletter
\providecommand \@ifxundefined [1]{%
 \@ifx{#1\undefined}
}%
\providecommand \@ifnum [1]{%
 \ifnum #1\expandafter \@firstoftwo
 \else \expandafter \@secondoftwo
 \fi
}%
\providecommand \@ifx [1]{%
 \ifx #1\expandafter \@firstoftwo
 \else \expandafter \@secondoftwo
 \fi
}%
\providecommand \natexlab [1]{#1}%
\providecommand \enquote  [1]{``#1''}%
\providecommand \bibnamefont  [1]{#1}%
\providecommand \bibfnamefont [1]{#1}%
\providecommand \citenamefont [1]{#1}%
\providecommand \href@noop [0]{\@secondoftwo}%
\providecommand \href [0]{\begingroup \@sanitize@url \@href}%
\providecommand \@href[1]{\@@startlink{#1}\@@href}%
\providecommand \@@href[1]{\endgroup#1\@@endlink}%
\providecommand \@sanitize@url [0]{\catcode `\\12\catcode `\$12\catcode
  `\&12\catcode `\#12\catcode `\^12\catcode `\_12\catcode `\%12\relax}%
\providecommand \@@startlink[1]{}%
\providecommand \@@endlink[0]{}%
\providecommand \url  [0]{\begingroup\@sanitize@url \@url }%
\providecommand \@url [1]{\endgroup\@href {#1}{\urlprefix }}%
\providecommand \urlprefix  [0]{URL }%
\providecommand \Eprint [0]{\href }%
\providecommand \doibase [0]{http://dx.doi.org/}%
\providecommand \selectlanguage [0]{\@gobble}%
\providecommand \bibinfo  [0]{\@secondoftwo}%
\providecommand \bibfield  [0]{\@secondoftwo}%
\providecommand \translation [1]{[#1]}%
\providecommand \BibitemOpen [0]{}%
\providecommand \bibitemStop [0]{}%
\providecommand \bibitemNoStop [0]{.\EOS\space}%
\providecommand \EOS [0]{\spacefactor3000\relax}%
\providecommand \BibitemShut  [1]{\csname bibitem#1\endcsname}%
\let\auto@bib@innerbib\@empty
\bibitem [{\citenamefont {Wolter}\ \emph {et~al.}(2015)\citenamefont {Wolter},
  \citenamefont {Pullen}, \citenamefont {Baudisch}, \citenamefont {Sclafani},
  \citenamefont {Hemmer}, \citenamefont {Senftleben}, \citenamefont
  {Schr\"oter}, \citenamefont {Ullrich}, \citenamefont {Moshammer},\ and\
  \citenamefont {Biegert}}]{Wolter2015}%
  \BibitemOpen
  \bibfield  {author} {\bibinfo {author} {\bibfnamefont {B.}~\bibnamefont
  {Wolter}}, \bibinfo {author} {\bibfnamefont {M.~G.}\ \bibnamefont {Pullen}},
  \bibinfo {author} {\bibfnamefont {M.}~\bibnamefont {Baudisch}}, \bibinfo
  {author} {\bibfnamefont {M.}~\bibnamefont {Sclafani}}, \bibinfo {author}
  {\bibfnamefont {M.}~\bibnamefont {Hemmer}}, \bibinfo {author} {\bibfnamefont
  {A.}~\bibnamefont {Senftleben}}, \bibinfo {author} {\bibfnamefont {C.~D.}\
  \bibnamefont {Schr\"oter}}, \bibinfo {author} {\bibfnamefont
  {J.}~\bibnamefont {Ullrich}}, \bibinfo {author} {\bibfnamefont
  {R.}~\bibnamefont {Moshammer}}, \ and\ \bibinfo {author} {\bibfnamefont
  {J.}~\bibnamefont {Biegert}},\ }\href {\doibase 10.1103/PhysRevX.5.021034}
  {\bibfield  {journal} {\bibinfo  {journal} {Phys. Rev. X}\ }\textbf {\bibinfo
  {volume} {5}},\ \bibinfo {pages} {021034} (\bibinfo {year}
  {2015})}\BibitemShut {NoStop}%
\bibitem [{\citenamefont {Keldysh}(1964)}]{Keldysh1964}%
  \BibitemOpen
  \bibfield  {author} {\bibinfo {author} {\bibfnamefont {L.}~\bibnamefont
  {Keldysh}},\ }\href@noop {} {\bibfield  {journal} {\bibinfo  {journal} {Zh.
  Eksp. Teor. Fiz.}\ }\textbf {\bibinfo {volume} {47}},\ \bibinfo {pages}
  {1945} (\bibinfo {year} {1964})},\ \bibinfo {note} {[Sov. Phys. JETP 20, 1307
  (1965)]}\BibitemShut {NoStop}%
\bibitem [{\citenamefont {Faisal}(1973)}]{Faisal1973}%
  \BibitemOpen
  \bibfield  {author} {\bibinfo {author} {\bibfnamefont {F.~H.~M.}\
  \bibnamefont {Faisal}},\ }\href {http://stacks.iop.org/0022-3700/6/i=4/a=011}
  {\bibfield  {journal} {\bibinfo  {journal} {Journal of Physics B: Atomic and
  Molecular Physics}\ }\textbf {\bibinfo {volume} {6}},\ \bibinfo {pages} {L89}
  (\bibinfo {year} {1973})}\BibitemShut {NoStop}%
\bibitem [{\citenamefont {Reiss}(1980)}]{Reiss1980}%
  \BibitemOpen
  \bibfield  {author} {\bibinfo {author} {\bibfnamefont {H.~R.}\ \bibnamefont
  {Reiss}},\ }\href {\doibase 10.1103/PhysRevA.22.1786} {\bibfield  {journal}
  {\bibinfo  {journal} {Phys. Rev. A}\ }\textbf {\bibinfo {volume} {22}},\
  \bibinfo {pages} {1786} (\bibinfo {year} {1980})}\BibitemShut {NoStop}%
\bibitem [{\citenamefont {Popov}(2004)}]{Popov2004}%
  \BibitemOpen
  \bibfield  {author} {\bibinfo {author} {\bibfnamefont {V.~S.}\ \bibnamefont
  {Popov}},\ }\href {http://stacks.iop.org/1063-7869/47/i=9/a=R01} {\bibfield
  {journal} {\bibinfo  {journal} {Physics-Uspekhi}\ }\textbf {\bibinfo {volume}
  {47}},\ \bibinfo {pages} {855} (\bibinfo {year} {2004})}\BibitemShut
  {NoStop}%
\bibitem [{\citenamefont {Kopold}\ \emph {et~al.}(2000)\citenamefont {Kopold},
  \citenamefont {Becker},\ and\ \citenamefont {Kleber}}]{Kopold2000}%
  \BibitemOpen
  \bibfield  {author} {\bibinfo {author} {\bibfnamefont {R.}~\bibnamefont
  {Kopold}}, \bibinfo {author} {\bibfnamefont {W.}~\bibnamefont {Becker}}, \
  and\ \bibinfo {author} {\bibfnamefont {M.}~\bibnamefont {Kleber}},\ }\href
  {\doibase http://dx.doi.org/10.1016/S0030-4018(99)00521-0} {\bibfield
  {journal} {\bibinfo  {journal} {Optics Communications}\ }\textbf {\bibinfo
  {volume} {179}},\ \bibinfo {pages} {39 } (\bibinfo {year}
  {2000})}\BibitemShut {NoStop}%
\bibitem [{\citenamefont {Salières}\ \emph {et~al.}(2001)\citenamefont
  {Salières}, \citenamefont {Carré}, \citenamefont {Le~Déroff},
  \citenamefont {Grasbon}, \citenamefont {Paulus}, \citenamefont {Walther},
  \citenamefont {Kopold}, \citenamefont {Becker}, \citenamefont {Milošević},
  \citenamefont {Sanpera},\ and\ \citenamefont {Lewenstein}}]{Salieres2001}%
  \BibitemOpen
  \bibfield  {author} {\bibinfo {author} {\bibfnamefont {P.}~\bibnamefont
  {Salières}}, \bibinfo {author} {\bibfnamefont {B.}~\bibnamefont {Carré}},
  \bibinfo {author} {\bibfnamefont {L.}~\bibnamefont {Le~Déroff}}, \bibinfo
  {author} {\bibfnamefont {F.}~\bibnamefont {Grasbon}}, \bibinfo {author}
  {\bibfnamefont {G.~G.}\ \bibnamefont {Paulus}}, \bibinfo {author}
  {\bibfnamefont {H.}~\bibnamefont {Walther}}, \bibinfo {author} {\bibfnamefont
  {R.}~\bibnamefont {Kopold}}, \bibinfo {author} {\bibfnamefont
  {W.}~\bibnamefont {Becker}}, \bibinfo {author} {\bibfnamefont {D.~B.}\
  \bibnamefont {Milošević}}, \bibinfo {author} {\bibfnamefont
  {A.}~\bibnamefont {Sanpera}}, \ and\ \bibinfo {author} {\bibfnamefont
  {M.}~\bibnamefont {Lewenstein}},\ }\href {\doibase 10.1126/science.108836}
  {\bibfield  {journal} {\bibinfo  {journal} {Science}\ }\textbf {\bibinfo
  {volume} {292}},\ \bibinfo {pages} {902} (\bibinfo {year}
  {2001})}\BibitemShut {NoStop}%
\bibitem [{\citenamefont {Milo\v{s}evi\'{c}}\ \emph {et~al.}(2006)\citenamefont
  {Milo\v{s}evi\'{c}}, \citenamefont {Paulus}, \citenamefont {Bauer},\ and\
  \citenamefont {Becker}}]{Milosevic2006}%
  \BibitemOpen
  \bibfield  {author} {\bibinfo {author} {\bibfnamefont {D.~B.}\ \bibnamefont
  {Milo\v{s}evi\'{c}}}, \bibinfo {author} {\bibfnamefont {G.~G.}\ \bibnamefont
  {Paulus}}, \bibinfo {author} {\bibfnamefont {D.}~\bibnamefont {Bauer}}, \
  and\ \bibinfo {author} {\bibfnamefont {W.}~\bibnamefont {Becker}},\ }\href
  {http://stacks.iop.org/0953-4075/39/i=14/a=R01} {\bibfield  {journal}
  {\bibinfo  {journal} {Journal of Physics B: Atomic, Molecular and Optical
  Physics}\ }\textbf {\bibinfo {volume} {39}},\ \bibinfo {pages} {R203}
  (\bibinfo {year} {2006})}\BibitemShut {NoStop}%
\bibitem [{\citenamefont {Popruzhenko}(2014{\natexlab{a}})}]{Popruzhenko2014a}%
  \BibitemOpen
  \bibfield  {author} {\bibinfo {author} {\bibfnamefont {S.~V.}\ \bibnamefont
  {Popruzhenko}},\ }\href {http://stacks.iop.org/0953-4075/47/i=20/a=204001}
  {\bibfield  {journal} {\bibinfo  {journal} {Journal of Physics B: Atomic,
  Molecular and Optical Physics}\ }\textbf {\bibinfo {volume} {47}},\ \bibinfo
  {pages} {204001} (\bibinfo {year} {2014}{\natexlab{a}})}\BibitemShut
  {NoStop}%
\bibitem [{\citenamefont {Lohr}\ \emph {et~al.}(1997)\citenamefont {Lohr},
  \citenamefont {Kleber}, \citenamefont {Kopold},\ and\ \citenamefont
  {Becker}}]{Lohr97}%
  \BibitemOpen
  \bibfield  {author} {\bibinfo {author} {\bibfnamefont {A.}~\bibnamefont
  {Lohr}}, \bibinfo {author} {\bibfnamefont {M.}~\bibnamefont {Kleber}},
  \bibinfo {author} {\bibfnamefont {R.}~\bibnamefont {Kopold}}, \ and\ \bibinfo
  {author} {\bibfnamefont {W.}~\bibnamefont {Becker}},\ }\href {\doibase
  10.1103/PhysRevA.55.R4003} {\bibfield  {journal} {\bibinfo  {journal} {Phys.
  Rev. A}\ }\textbf {\bibinfo {volume} {55}},\ \bibinfo {pages} {R4003}
  (\bibinfo {year} {1997})}\BibitemShut {NoStop}%
\bibitem [{\citenamefont {Milo\v{s}evi\'{c}}\ and\ \citenamefont
  {Ehlotzky}(1998)}]{Milosevic1998}%
  \BibitemOpen
  \bibfield  {author} {\bibinfo {author} {\bibfnamefont {D.~B.}\ \bibnamefont
  {Milo\v{s}evi\'{c}}}\ and\ \bibinfo {author} {\bibfnamefont {F.}~\bibnamefont
  {Ehlotzky}},\ }\href {\doibase 10.1103/PhysRevA.58.3124} {\bibfield
  {journal} {\bibinfo  {journal} {Phys. Rev. A}\ }\textbf {\bibinfo {volume}
  {58}},\ \bibinfo {pages} {3124} (\bibinfo {year} {1998})}\BibitemShut
  {NoStop}%
\bibitem [{\citenamefont {M\"oller}\ \emph {et~al.}(2014)\citenamefont
  {M\"oller}, \citenamefont {Meyer}, \citenamefont {Sayler}, \citenamefont
  {Paulus}, \citenamefont {Kling}, \citenamefont {Schmidt}, \citenamefont
  {Becker},\ and\ \citenamefont {Milo\ifmmode \check{s}\else
  \v{s}\fi{}evi\ifmmode~\acute{c}\else \'{c}\fi{}}}]{Moeller14}%
  \BibitemOpen
  \bibfield  {author} {\bibinfo {author} {\bibfnamefont {M.}~\bibnamefont
  {M\"oller}}, \bibinfo {author} {\bibfnamefont {F.}~\bibnamefont {Meyer}},
  \bibinfo {author} {\bibfnamefont {A.~M.}\ \bibnamefont {Sayler}}, \bibinfo
  {author} {\bibfnamefont {G.~G.}\ \bibnamefont {Paulus}}, \bibinfo {author}
  {\bibfnamefont {M.~F.}\ \bibnamefont {Kling}}, \bibinfo {author}
  {\bibfnamefont {B.~E.}\ \bibnamefont {Schmidt}}, \bibinfo {author}
  {\bibfnamefont {W.}~\bibnamefont {Becker}}, \ and\ \bibinfo {author}
  {\bibfnamefont {D.~B.}\ \bibnamefont {Milo\ifmmode \check{s}\else
  \v{s}\fi{}evi\ifmmode~\acute{c}\else \'{c}\fi{}}},\ }\href {\doibase
  10.1103/PhysRevA.90.023412} {\bibfield  {journal} {\bibinfo  {journal} {Phys.
  Rev. A}\ }\textbf {\bibinfo {volume} {90}},\ \bibinfo {pages} {023412}
  (\bibinfo {year} {2014})}\BibitemShut {NoStop}%
\bibitem [{\citenamefont {Blaga}\ \emph {et~al.}(2009)\citenamefont {Blaga},
  \citenamefont {Catoire}, \citenamefont {Colosimo}, \citenamefont {Paulus},
  \citenamefont {Muller}, \citenamefont {Agostini},\ and\ \citenamefont
  {DiMauro}}]{Blaga2009}%
  \BibitemOpen
  \bibfield  {author} {\bibinfo {author} {\bibfnamefont {C.~I.}\ \bibnamefont
  {Blaga}}, \bibinfo {author} {\bibfnamefont {F.}~\bibnamefont {Catoire}},
  \bibinfo {author} {\bibfnamefont {P.}~\bibnamefont {Colosimo}}, \bibinfo
  {author} {\bibfnamefont {G.~G.}\ \bibnamefont {Paulus}}, \bibinfo {author}
  {\bibfnamefont {H.~G.}\ \bibnamefont {Muller}}, \bibinfo {author}
  {\bibfnamefont {P.}~\bibnamefont {Agostini}}, \ and\ \bibinfo {author}
  {\bibfnamefont {L.~F.}\ \bibnamefont {DiMauro}},\ }\href {\doibase
  10.1038/nphys1228} {\bibfield  {journal} {\bibinfo  {journal} {Nature
  Physics}\ }\textbf {\bibinfo {volume} {5}},\ \bibinfo {pages} {335} (\bibinfo
  {year} {2009})}\BibitemShut {NoStop}%
\bibitem [{\citenamefont {Faisal}(2009)}]{Faisal2009}%
  \BibitemOpen
  \bibfield  {author} {\bibinfo {author} {\bibfnamefont {F.~H.~M.}\
  \bibnamefont {Faisal}},\ }\href {\doibase 10.1038/nphys1264} {\bibfield
  {journal} {\bibinfo  {journal} {Nature Physics}\ }\textbf {\bibinfo {volume}
  {5}},\ \bibinfo {pages} {319} (\bibinfo {year} {2009})}\BibitemShut {NoStop}%
\bibitem [{\citenamefont {Quan}\ \emph {et~al.}(2009)\citenamefont {Quan},
  \citenamefont {Lin}, \citenamefont {Wu}, \citenamefont {Kang}, \citenamefont
  {Liu}, \citenamefont {Liu}, \citenamefont {Chen}, \citenamefont {Liu},
  \citenamefont {He}, \citenamefont {Chen}, \citenamefont {Xiong},
  \citenamefont {Guo}, \citenamefont {Xu}, \citenamefont {Fu}, \citenamefont
  {Cheng},\ and\ \citenamefont {Xu}}]{Quan2009}%
  \BibitemOpen
  \bibfield  {author} {\bibinfo {author} {\bibfnamefont {W.}~\bibnamefont
  {Quan}}, \bibinfo {author} {\bibfnamefont {Z.}~\bibnamefont {Lin}}, \bibinfo
  {author} {\bibfnamefont {M.}~\bibnamefont {Wu}}, \bibinfo {author}
  {\bibfnamefont {H.}~\bibnamefont {Kang}}, \bibinfo {author} {\bibfnamefont
  {H.}~\bibnamefont {Liu}}, \bibinfo {author} {\bibfnamefont {X.}~\bibnamefont
  {Liu}}, \bibinfo {author} {\bibfnamefont {J.}~\bibnamefont {Chen}}, \bibinfo
  {author} {\bibfnamefont {J.}~\bibnamefont {Liu}}, \bibinfo {author}
  {\bibfnamefont {X.~T.}\ \bibnamefont {He}}, \bibinfo {author} {\bibfnamefont
  {S.~G.}\ \bibnamefont {Chen}}, \bibinfo {author} {\bibfnamefont
  {H.}~\bibnamefont {Xiong}}, \bibinfo {author} {\bibfnamefont
  {L.}~\bibnamefont {Guo}}, \bibinfo {author} {\bibfnamefont {H.}~\bibnamefont
  {Xu}}, \bibinfo {author} {\bibfnamefont {Y.}~\bibnamefont {Fu}}, \bibinfo
  {author} {\bibfnamefont {Y.}~\bibnamefont {Cheng}}, \ and\ \bibinfo {author}
  {\bibfnamefont {Z.~Z.}\ \bibnamefont {Xu}},\ }\href {\doibase
  10.1103/PhysRevLett.103.093001} {\bibfield  {journal} {\bibinfo  {journal}
  {Phys. Rev. Lett.}\ }\textbf {\bibinfo {volume} {103}},\ \bibinfo {pages}
  {093001} (\bibinfo {year} {2009})}\BibitemShut {NoStop}%
\bibitem [{\citenamefont {Wu}\ \emph {et~al.}(2012)\citenamefont {Wu},
  \citenamefont {Yang}, \citenamefont {Liu}, \citenamefont {Gong},
  \citenamefont {Wu}, \citenamefont {Liu}, \citenamefont {Hao}, \citenamefont
  {Li}, \citenamefont {He},\ and\ \citenamefont {Chen}}]{Wu2012}%
  \BibitemOpen
  \bibfield  {author} {\bibinfo {author} {\bibfnamefont {C.~Y.}\ \bibnamefont
  {Wu}}, \bibinfo {author} {\bibfnamefont {Y.~D.}\ \bibnamefont {Yang}},
  \bibinfo {author} {\bibfnamefont {Y.~Q.}\ \bibnamefont {Liu}}, \bibinfo
  {author} {\bibfnamefont {Q.~H.}\ \bibnamefont {Gong}}, \bibinfo {author}
  {\bibfnamefont {M.}~\bibnamefont {Wu}}, \bibinfo {author} {\bibfnamefont
  {X.}~\bibnamefont {Liu}}, \bibinfo {author} {\bibfnamefont {X.~L.}\
  \bibnamefont {Hao}}, \bibinfo {author} {\bibfnamefont {W.~D.}\ \bibnamefont
  {Li}}, \bibinfo {author} {\bibfnamefont {X.~T.}\ \bibnamefont {He}}, \ and\
  \bibinfo {author} {\bibfnamefont {J.}~\bibnamefont {Chen}},\ }\href {\doibase
  10.1103/PhysRevLett.109.043001} {\bibfield  {journal} {\bibinfo  {journal}
  {Phys. Rev. Lett.}\ }\textbf {\bibinfo {volume} {109}},\ \bibinfo {pages}
  {043001} (\bibinfo {year} {2012})}\BibitemShut {NoStop}%
\bibitem [{\citenamefont {Dura}\ \emph {et~al.}(2013)\citenamefont {Dura},
  \citenamefont {Camus}, \citenamefont {Thai}, \citenamefont {Britz},
  \citenamefont {Hemmer}, \citenamefont {Baudisch}, \citenamefont {Senftleben},
  \citenamefont {Schröter}, \citenamefont {Ullrich}, \citenamefont
  {Moshammer},\ and\ \citenamefont {Biegert}}]{Dura2013}%
  \BibitemOpen
  \bibfield  {author} {\bibinfo {author} {\bibfnamefont {J.}~\bibnamefont
  {Dura}}, \bibinfo {author} {\bibfnamefont {N.}~\bibnamefont {Camus}},
  \bibinfo {author} {\bibfnamefont {A.}~\bibnamefont {Thai}}, \bibinfo {author}
  {\bibfnamefont {A.}~\bibnamefont {Britz}}, \bibinfo {author} {\bibfnamefont
  {M.}~\bibnamefont {Hemmer}}, \bibinfo {author} {\bibfnamefont
  {M.}~\bibnamefont {Baudisch}}, \bibinfo {author} {\bibfnamefont
  {A.}~\bibnamefont {Senftleben}}, \bibinfo {author} {\bibfnamefont {C.~D.}\
  \bibnamefont {Schröter}}, \bibinfo {author} {\bibfnamefont {J.}~\bibnamefont
  {Ullrich}}, \bibinfo {author} {\bibfnamefont {R.}~\bibnamefont {Moshammer}},
  \ and\ \bibinfo {author} {\bibfnamefont {J.}~\bibnamefont {Biegert}},\ }\href
  {\doibase 10.1038/srep02675} {\bibfield  {journal} {\bibinfo  {journal}
  {Scientific Reports}\ }\textbf {\bibinfo {volume} {3}},\ \bibinfo {pages}
  {2675} (\bibinfo {year} {2013})}\BibitemShut {NoStop}%
\bibitem [{\citenamefont {Liu}\ and\ \citenamefont
  {Hatsagortsyan}(2010)}]{LiuHatsag10}%
  \BibitemOpen
  \bibfield  {author} {\bibinfo {author} {\bibfnamefont {C.}~\bibnamefont
  {Liu}}\ and\ \bibinfo {author} {\bibfnamefont {K.~Z.}\ \bibnamefont
  {Hatsagortsyan}},\ }\href {\doibase 10.1103/PhysRevLett.105.113003}
  {\bibfield  {journal} {\bibinfo  {journal} {Phys. Rev. Lett.}\ }\textbf
  {\bibinfo {volume} {105}},\ \bibinfo {pages} {113003} (\bibinfo {year}
  {2010})}\BibitemShut {NoStop}%
\bibitem [{\citenamefont {K\"astner}\ \emph {et~al.}(2012)\citenamefont
  {K\"astner}, \citenamefont {Saalmann},\ and\ \citenamefont
  {Rost}}]{Kaestner12}%
  \BibitemOpen
  \bibfield  {author} {\bibinfo {author} {\bibfnamefont {A.}~\bibnamefont
  {K\"astner}}, \bibinfo {author} {\bibfnamefont {U.}~\bibnamefont {Saalmann}},
  \ and\ \bibinfo {author} {\bibfnamefont {J.~M.}\ \bibnamefont {Rost}},\
  }\href {\doibase 10.1103/PhysRevLett.108.033201} {\bibfield  {journal}
  {\bibinfo  {journal} {Phys. Rev. Lett.}\ }\textbf {\bibinfo {volume} {108}},\
  \bibinfo {pages} {033201} (\bibinfo {year} {2012})}\BibitemShut {NoStop}%
\bibitem [{\citenamefont {Milo\v{s}evi\'{c}}(2014)}]{Milosevic2014}%
  \BibitemOpen
  \bibfield  {author} {\bibinfo {author} {\bibfnamefont {D.~B.}\ \bibnamefont
  {Milo\v{s}evi\'{c}}},\ }\href {\doibase 10.1103/PhysRevA.90.063423}
  {\bibfield  {journal} {\bibinfo  {journal} {Phys. Rev. A}\ }\textbf {\bibinfo
  {volume} {90}},\ \bibinfo {pages} {063423} (\bibinfo {year}
  {2014})}\BibitemShut {NoStop}%
\bibitem [{\citenamefont {Arb\'o}\ \emph {et~al.}(2010)\citenamefont {Arb\'o},
  \citenamefont {Ishikawa}, \citenamefont {Schiessl}, \citenamefont {Persson},\
  and\ \citenamefont {Burgd\"orfer}}]{Arbo10}%
  \BibitemOpen
  \bibfield  {author} {\bibinfo {author} {\bibfnamefont {D.~G.}\ \bibnamefont
  {Arb\'o}}, \bibinfo {author} {\bibfnamefont {K.~L.}\ \bibnamefont
  {Ishikawa}}, \bibinfo {author} {\bibfnamefont {K.}~\bibnamefont {Schiessl}},
  \bibinfo {author} {\bibfnamefont {E.}~\bibnamefont {Persson}}, \ and\
  \bibinfo {author} {\bibfnamefont {J.}~\bibnamefont {Burgd\"orfer}},\ }\href
  {\doibase 10.1103/PhysRevA.81.021403} {\bibfield  {journal} {\bibinfo
  {journal} {Phys. Rev. A}\ }\textbf {\bibinfo {volume} {81}},\ \bibinfo
  {pages} {021403} (\bibinfo {year} {2010})}\BibitemShut {NoStop}%
\bibitem [{\citenamefont {Yan}\ \emph {et~al.}(2010)\citenamefont {Yan},
  \citenamefont {Popruzhenko}, \citenamefont {Vrakking},\ and\ \citenamefont
  {Bauer}}]{Yan2010}%
  \BibitemOpen
  \bibfield  {author} {\bibinfo {author} {\bibfnamefont {T.-M.}\ \bibnamefont
  {Yan}}, \bibinfo {author} {\bibfnamefont {S.~V.}\ \bibnamefont
  {Popruzhenko}}, \bibinfo {author} {\bibfnamefont {M.~J.~J.}\ \bibnamefont
  {Vrakking}}, \ and\ \bibinfo {author} {\bibfnamefont {D.}~\bibnamefont
  {Bauer}},\ }\href {\doibase 10.1103/PhysRevLett.105.253002} {\bibfield
  {journal} {\bibinfo  {journal} {Phys. Rev. Lett.}\ }\textbf {\bibinfo
  {volume} {105}},\ \bibinfo {pages} {253002} (\bibinfo {year}
  {2010})}\BibitemShut {NoStop}%
\bibitem [{\citenamefont {Popruzhenko}(2014{\natexlab{b}})}]{Popruzhenko2014}%
  \BibitemOpen
  \bibfield  {author} {\bibinfo {author} {\bibfnamefont {S.}~\bibnamefont
  {Popruzhenko}},\ }\href {\doibase 10.1134/S1063776114040062} {\bibfield
  {journal} {\bibinfo  {journal} {Journal of Experimental and Theoretical
  Physics}\ }\textbf {\bibinfo {volume} {118}},\ \bibinfo {pages} {580}
  (\bibinfo {year} {2014}{\natexlab{b}})}\BibitemShut {NoStop}%
\bibitem [{\citenamefont {Pisanty}\ and\ \citenamefont
  {Ivanov}(2016)}]{Pisanty2016}%
  \BibitemOpen
  \bibfield  {author} {\bibinfo {author} {\bibfnamefont {E.}~\bibnamefont
  {Pisanty}}\ and\ \bibinfo {author} {\bibfnamefont {M.}~\bibnamefont
  {Ivanov}},\ }\href {\doibase 10.1103/PhysRevA.93.043408} {\bibfield
  {journal} {\bibinfo  {journal} {Phys. Rev. A}\ }\textbf {\bibinfo {volume}
  {93}},\ \bibinfo {pages} {043408} (\bibinfo {year} {2016})}\BibitemShut
  {NoStop}%
\bibitem [{\citenamefont {Huismans}\ \emph {et~al.}(2011)\citenamefont
  {Huismans}, \citenamefont {Rouzée}, \citenamefont {Gijsbertsen},
  \citenamefont {Jungmann}, \citenamefont {Smolkowska}, \citenamefont {Logman},
  \citenamefont {Lépine}, \citenamefont {Cauchy}, \citenamefont {Zamith},
  \citenamefont {Marchenko}, \citenamefont {Bakker}, \citenamefont {Berden},
  \citenamefont {Redlich}, \citenamefont {van~der Meer}, \citenamefont
  {Muller}, \citenamefont {Vermin}, \citenamefont {Schafer}, \citenamefont
  {Spanner}, \citenamefont {Ivanov}, \citenamefont {Smirnova}, \citenamefont
  {Bauer}, \citenamefont {Popruzhenko},\ and\ \citenamefont
  {Vrakking}}]{Huismans2011}%
  \BibitemOpen
  \bibfield  {author} {\bibinfo {author} {\bibfnamefont {Y.}~\bibnamefont
  {Huismans}}, \bibinfo {author} {\bibfnamefont {A.}~\bibnamefont {Rouzée}},
  \bibinfo {author} {\bibfnamefont {A.}~\bibnamefont {Gijsbertsen}}, \bibinfo
  {author} {\bibfnamefont {J.~H.}\ \bibnamefont {Jungmann}}, \bibinfo {author}
  {\bibfnamefont {A.~S.}\ \bibnamefont {Smolkowska}}, \bibinfo {author}
  {\bibfnamefont {P.~S. W.~M.}\ \bibnamefont {Logman}}, \bibinfo {author}
  {\bibfnamefont {F.}~\bibnamefont {Lépine}}, \bibinfo {author} {\bibfnamefont
  {C.}~\bibnamefont {Cauchy}}, \bibinfo {author} {\bibfnamefont
  {S.}~\bibnamefont {Zamith}}, \bibinfo {author} {\bibfnamefont
  {T.}~\bibnamefont {Marchenko}}, \bibinfo {author} {\bibfnamefont {J.~M.}\
  \bibnamefont {Bakker}}, \bibinfo {author} {\bibfnamefont {G.}~\bibnamefont
  {Berden}}, \bibinfo {author} {\bibfnamefont {B.}~\bibnamefont {Redlich}},
  \bibinfo {author} {\bibfnamefont {A.~F.~G.}\ \bibnamefont {van~der Meer}},
  \bibinfo {author} {\bibfnamefont {H.~G.}\ \bibnamefont {Muller}}, \bibinfo
  {author} {\bibfnamefont {W.}~\bibnamefont {Vermin}}, \bibinfo {author}
  {\bibfnamefont {K.~J.}\ \bibnamefont {Schafer}}, \bibinfo {author}
  {\bibfnamefont {M.}~\bibnamefont {Spanner}}, \bibinfo {author} {\bibfnamefont
  {M.~Y.}\ \bibnamefont {Ivanov}}, \bibinfo {author} {\bibfnamefont
  {O.}~\bibnamefont {Smirnova}}, \bibinfo {author} {\bibfnamefont
  {D.}~\bibnamefont {Bauer}}, \bibinfo {author} {\bibfnamefont {S.~V.}\
  \bibnamefont {Popruzhenko}}, \ and\ \bibinfo {author} {\bibfnamefont
  {M.~J.~J.}\ \bibnamefont {Vrakking}},\ }\href {\doibase
  10.1126/science.1198450} {\bibfield  {journal} {\bibinfo  {journal}
  {Science}\ }\textbf {\bibinfo {volume} {331}},\ \bibinfo {pages} {61}
  (\bibinfo {year} {2011})}\BibitemShut {NoStop}%
\bibitem [{\citenamefont {Corkum}\ \emph {et~al.}(1989)\citenamefont {Corkum},
  \citenamefont {Burnett},\ and\ \citenamefont {Brunel}}]{CorkumSMT}%
  \BibitemOpen
  \bibfield  {author} {\bibinfo {author} {\bibfnamefont {P.~B.}\ \bibnamefont
  {Corkum}}, \bibinfo {author} {\bibfnamefont {N.~H.}\ \bibnamefont {Burnett}},
  \ and\ \bibinfo {author} {\bibfnamefont {F.}~\bibnamefont {Brunel}},\ }\href
  {\doibase 10.1103/PhysRevLett.62.1259} {\bibfield  {journal} {\bibinfo
  {journal} {Phys. Rev. Lett.}\ }\textbf {\bibinfo {volume} {62}},\ \bibinfo
  {pages} {1259} (\bibinfo {year} {1989})}\BibitemShut {NoStop}%
\bibitem [{Note1()}]{Note1}%
  \BibitemOpen
  \bibinfo {note} {The scaling in SI units is $\tau =\omega t$ and $\protect
  \bm {\mathaccent "7016\relax {r}} = \protect \sqrt {m\omega /{\mathchar
  '26\mkern -9muh}}~{\protect \bm {r}}$.}\BibitemShut {Stop}%
\bibitem [{Note2()}]{Note2}%
  \BibitemOpen
  \bibinfo {note} {In general, PES also depend on the other quantum numbers
  such as orbital angular momentum $l$ and magnetic quantum number $m$ but this
  dependence is not related to the yield enhancement discussed in this
  work.}\BibitemShut {Stop}%
\bibitem [{Note3()}]{Note3}%
  \BibitemOpen
  \bibinfo {note} {We found in Fig.~1 of Ref.~\cite {Milosevic1998} a flat PES
  plateau up to $2U_{\protect \mathrm {p}}$, although without discussion. The
  PES there was calculated using a Coulomb-Volkov ansatz.}\BibitemShut {Stop}%
\bibitem [{\citenamefont {Popruzhenko}\ and\ \citenamefont
  {Bauer}(2008)}]{Popruzhenko2008}%
  \BibitemOpen
  \bibfield  {author} {\bibinfo {author} {\bibfnamefont {S.}~\bibnamefont
  {Popruzhenko}}\ and\ \bibinfo {author} {\bibfnamefont {D.}~\bibnamefont
  {Bauer}},\ }\href {\doibase 10.1080/09500340802161881} {\bibfield  {journal}
  {\bibinfo  {journal} {Journal of Modern Optics}\ }\textbf {\bibinfo {volume}
  {55}},\ \bibinfo {pages} {2573} (\bibinfo {year} {2008})}\BibitemShut
  {NoStop}%
\bibitem [{\citenamefont {Torlina}\ and\ \citenamefont
  {Smirnova}(2012)}]{Torlina2012}%
  \BibitemOpen
  \bibfield  {author} {\bibinfo {author} {\bibfnamefont {L.}~\bibnamefont
  {Torlina}}\ and\ \bibinfo {author} {\bibfnamefont {O.}~\bibnamefont
  {Smirnova}},\ }\href {\doibase 10.1103/PhysRevA.86.043408} {\bibfield
  {journal} {\bibinfo  {journal} {Phys. Rev. A}\ }\textbf {\bibinfo {volume}
  {86}},\ \bibinfo {pages} {043408} (\bibinfo {year} {2012})}\BibitemShut
  {NoStop}%
\bibitem [{\citenamefont {Kaushal}\ and\ \citenamefont
  {Smirnova}(2013)}]{Kaushal2013}%
  \BibitemOpen
  \bibfield  {author} {\bibinfo {author} {\bibfnamefont {J.}~\bibnamefont
  {Kaushal}}\ and\ \bibinfo {author} {\bibfnamefont {O.}~\bibnamefont
  {Smirnova}},\ }\href {\doibase 10.1103/PhysRevA.88.013421} {\bibfield
  {journal} {\bibinfo  {journal} {Phys. Rev. A}\ }\textbf {\bibinfo {volume}
  {88}},\ \bibinfo {pages} {013421} (\bibinfo {year} {2013})}\BibitemShut
  {NoStop}%
\end{thebibliography}%


\end{document}